# Wurtzite AlScN/AlN Superlattice Ferroelectrics Enable Endurance Beyond $10^{10}$ Cycles

Ruiqing Wang[1]†, Feng Zhu[3]†, Haoji Qian[2], Jiuren Zhou[1,2]*, Wenxin Sun[1], Siying Zheng[1,2], Jiajia Chen[1,2], Bochang Li[1,2], Yan Liu[1,2], Peng Zhou[4], Yue Hao[1] & Genquan Han[1,2]*

[1] *Faculty of Integrated Circuit, Xidian University, Xi'an, China.*

[2] *Hangzhou Institute of Technology, Xidian University, Hangzhou, China.*

[3] *TRACE EM Unit and Department of Materials Science and Engineering, City University of Hong Kong, Hong Kong, China.*

[4] *State Key Laboratory of Integrated Chips and Systems, College of Integrated Circuits and Micro-Nano Electronics, Frontier Institute of Chip and System, Zhangjiang Fudan International Innovation Center, Fudan University, Shanghai, China.*

*Corresponding authors. Email: zhoujiuren@163.com; gqhan@xidian.edu.cn;

† Ruiqing Wang and Feng Zhu contributed equally to this work.

**Abstract**

Wurtzite ferroelectrics are rapidly emerging as a promising material class for next-generation non-volatile memory technologies, owing to their large remanent polarization, intrinsically ordered three-dimensional crystal structure and full compatibility with CMOS processes and back-end-of-line (BEOL) integration. However, their practical implementation remains critically constrained by a severe endurance bottleneck: under conditions where the remanent polarization ($2P_r$) reaches or exceeds 200 μC/cm$^2$, devices typically undergo catastrophic failure before reaching $10^8$ cycles. Here, we report a vacancy-confining superlattice strategy that addresses this limitation, achieving reliable ferroelectric switching beyond $10^{10}$ cycles while preserving saturated polarization ($2P_r \geq 200$ μC/cm$^2$). This is achieved by embedding periodic ultrathin AlN layers within AlScN films, forming wurtzite AlScN/AlN superlattices, in conjunction with a dynamic recovery protocol that actively stabilizes the defect landscape throughout repeated cycling. Atomic-resolution imaging and EELS spectrum imaging technique, supported by first-principles calculations, reveal a self-regulated defect topology in which nitrogen vacancies are spatially confined by heterostructure energy barriers and dynamically re-trapped into energetically favorable lattice sites. This dual spatial-energetic confinement mechanism effectively inhibits both long-range percolative migration and local defect clustering, enabling such an ultrahigh endurance exceeding $10^{10}$ cycles and limiting polarization degradation to below 3% after $10^9$ cycles.

These findings establish nitrogen vacancy topology stabilization as a foundational design principle for reliable operation of wurtzite ferroelectrics, providing a scalable and CMOS-compatible platform for future high-endurance ferroelectric memory technologies.

**Introduction**

Ferroelectric memory technologies are fundamentally reshaping non-volatile data storage by replacing conventional charge-based mechanisms with polarization control at the atomic scale[1,2]. This paradigm shift offers substantial advantages, including reduced read/write latency, improved energy efficiency, enhanced endurance and superior scalability[3-5]. Collectively, these attributes position ferroelectric memories as a compelling solution to address the rapidly escalating global data demand, which has already surpassed the zettabyte scale[6-8]. Such advances have been propelled by the emergence of CMOS-compatible ferroelectrics, most notably fluorite-structured hafnium oxides and wurtzite-phase nitrides, which have enabled the integration of ferroelectric functionality into mainstream semiconductor platforms[9-11].

Among these materials, wurtzite ferroelectrics have emerged as a promising platform for post-Moore high-density non-volatile memory technologies[12]. Their highly ordered three-dimensional lattice mitigates device-to-device variability, allowing for aggressive footprint scaling and large-scale integration. The coexistence of large spontaneous polarization ($2P_r \geq 200\ \mu C/cm^2$) and strong coercive fields ($E_c \geq 4\ MV/cm$) supports multi-level storage, while a low thermal budget processing window below 400 °C ensures full compatibility with back-end-of-line (BEOL) integration[13-15]. These advantages facilitate the transition from planar architectures to three-dimensional ferroelectric memory schemes[16]. Recent progress in film thickness scaling, silicon process integration and prototype demonstrations has further accelerated the trajectory toward practical deployment[17-23]. However, despite these achievements, wurtzite ferroelectrics still face a fundamental endurance limitation. Under saturated switching conditions ($2P_r \geq 200\ \mu C/cm^2$), catastrophic failure typically occurs before $10^8$ cycles due to irreversible hard breakdown[24-30].

In this work, we demonstrate a superlattice-based strategy that enables endurance in wurtzite ferroelectrics to exceed $10^{10}$ switching cycles, two orders of magnitude higher than previously reported limits, while maintaining saturated polarization ($2P_r \geq 200$ μC/cm$^2$). Polarization degradation remains below 3% after $10^9$ cycles, with endurance of over $4.6 \times 10^9$ cycles sustained at 300 K. Beyond endurance enhancement, our findings establish a conceptual foundation for vacancy-confining design and define a broadly applicable defect-engineering framework for enabling reliable ferroelectric operation in CMOS-compatible wurtzite systems, paving the way for next-generation non-volatile memory technologies.

**Vacancy-driven breakdown mechanism in bulk wurtzite ferroelectrics**

Identifying the fundamental mechanisms that limit endurance in bulk wurtzite ferroelectrics is crucial for unlocking their full potential in next-generation non-volatile memory technologies. Using aluminum scandium nitride (AlScN) as a model system representative of such material class, we examined the structural and electrical evolution of AlScN films subjected to $2 \times 10^6$ polarization switching cycles, immediately preceding catastrophic breakdown. High-resolution transmission electron microscopy (HRTEM) and selected-area electron diffraction (SAED) reveal that the long-range ferroelectric phase remains intact, with the c-axis consistently aligned perpendicular to the growth interface[31] (Figure 1a). However, vertically aligned grain boundaries extending from the top electrode (TE) to the bottom electrode (BE) become increasingly prominent after cycling, coinciding with the onset of hard breakdown[32-34].

Electrical characterization reveals a pronounced decoupling between ferroelectricity and leakage characteristics under stress. As shown in Figure 1b, the $P_r$ decreases by only ~8% after cycling, indicating that the ferroelectric lattice remains largely preserved. In contrast, the leakage current increases sharply, eventually exceeding a critical threshold and inducing breakdown, as evidenced by the abrupt collapse of $P_r$ in Figure 1c. Statistical analysis across ten devices confirms this trend, showing a ninefold increase in leakage current density—from 0.056 A/cm² to 0.49 A/cm² at 3 MV/cm, after $2\times10^6$ cycling (Figure 1d). These results indicate that endurance failure in bulk AlScN is not primarily driven by ferroelectricity loss, but rather by the emergence of electrically active conduction pathways formed under prolonged electrical stress[35-36].

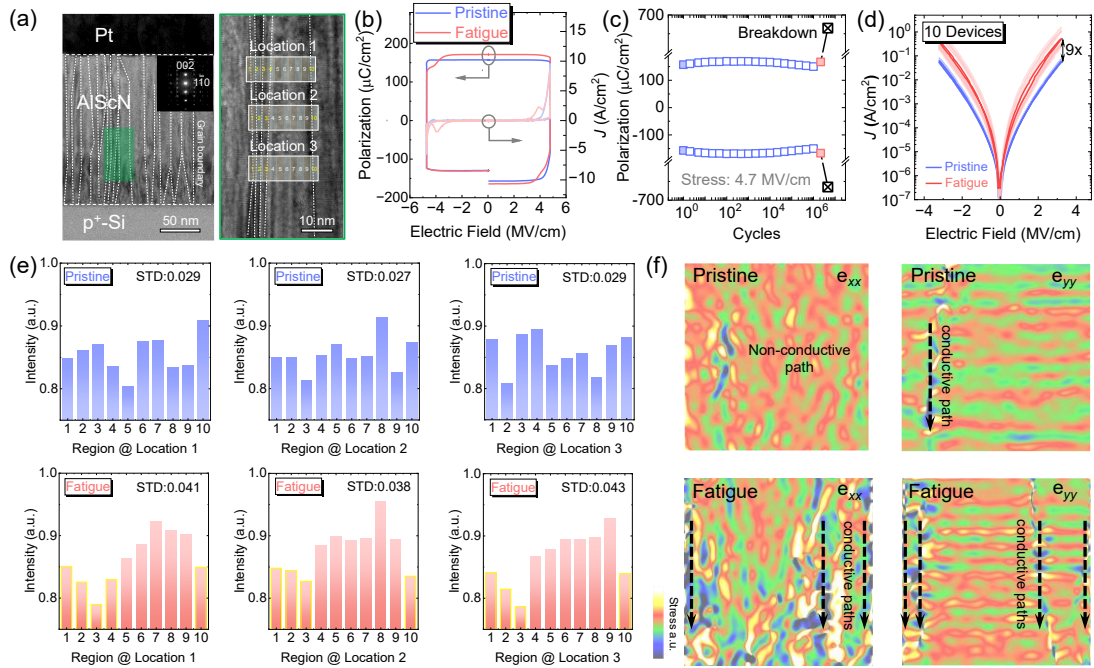

**Figure 1 Structural and electrical evolution preceding breakdown in bulk AlScN after $2\times10^6$ switching cycles. a,** Cross-sectional HRTEM reveals intact ferroelectric phase and vertically aligned grain boundaries spanning the film thickness. **b-d,** Electrical measurements show limited $P_r$ degradation but a sharp rise in leakage current, culminating in breakdown. **e-f,** Complementary EELS and strain mapping together reveal lateral clustering and vertical percolation of nitrogen vacancies, forming conductive channels that directly link vacancy redistribution to dielectric failure and highlighting the critical role of vacancy confinement for endurance improvement.

To uncover the microscopic origin of this leakage evolution, electron energy-loss spectroscopy (EELS) was performed on both pristine and fatigued AlScN samples. The impact of electrical cycling on nitrogen vacancy ($V_N$) distribution was assessed via intensity variations in the nitrogen K-edge (centered at 403.1 eV), using the scandium L-edge (407.5 eV) as an internal reference due to its fixed stoichiometry (Extended Data Figure 1)[37-40]. In the pristine state, the intensity for ten regions is relatively uniform and random across three locations, suggesting a homogeneous $V_N$ distribution (Figure 1e). After cycling, however, the intensity landscape becomes distinctly inhomogeneous: signal intensity decreases near grain boundaries and increases in the adjacent matrix, indicating lateral migration and clustering of $V_N$ toward the boundaries. This redistribution is quantitatively captured by the increase in the standard deviation (STD) of intensity from 0.029 (0.027~0.029) to 0.041 (0.038~0.043). These patterns are consistently observed across multiple regions and locations, pointing to a stress-driven vacancy percolation process. Complementary geometric phase analysis (GPA), specializing in strain mapping, further reinforces this interpretation[41]. As shown in Figure 1f, after $2\times10^6$ cycles, nitrogen vacancies accumulate and coalesce into conductive paths that span the entire film thickness. These vertically aligned, vacancy-rich channels provide a direct pathway for leakage current and form the physical basis for the abrupt transition to hard breakdown. Together, these results establish that endurance failure in bulk wurtzite ferroelectrics is governed by a vacancy-mediated conduction mechanism, initiated by lateral vacancy clustering and exacerbated by vertical percolation. Suppressing both transport pathways is thus critical for achieving stable and reliable operation in wurtzite-based ferroelectric memory devices.

**Wurtzite ferroelectric AlScN/AlN superlattice enables endurance breakthrough beyond $10^{10}$ cycles**

To suppress vertical nitrogen vacancy transport, a key driver of hard breakdown in wurtzite ferroelectrics, we first examined the site-specific energetics of vacancy formation in aluminum scandium nitride (AlScN) via first-principles density functional theory (DFT). Eight representative nitrogen sites were evaluated to reflect variations in local bonding environments (Figure 2a). The calculated formation energies ($E_{form}$) exhibit a bimodal distribution: ~3.8 eV for nitrogen atoms coordinated with scandium, and ~4.7 eV for those bonded solely to aluminum. This asymmetry originates from local coordination environments and indicates that Sc-rich regions, while essential for reducing the coercive field ($E_c$), are intrinsically more susceptible to vacancy formation. In conventional AlScN films with a homogeneous Sc distribution, these low-$E_{form}$ sites inevitably create contiguous pathways that promote long-range vacancy percolation, ultimately limiting endurance. To counteract this intrinsic vulnerability, we introduced periodic aluminum nitride (AlN) interlayers, free of scandium, into the AlScN matrix, forming a wurtzite AlScN/AlN superlattice that spatially modulates the vacancy formation energetics (Figure 2b). DFT calculations confirm that nitrogen atoms within the AlN layers, particularly those lacking Sc coordination, possess significantly elevated $E_{form}$, thereby suppressing vacancy formation. Moreover, the migration barrier for nitrogen vacancies across the AlN interlayer (e.g., $V_{N-A} \rightarrow V_{N-C}$) is substantially higher than that within the AlScN domain (e.g., $V_{N-I} \rightarrow V_{N-J}$). All these results suggest that AlN interlayer serves as a key energetic barrier to $V_N$ transport that can impede through-thickness vacancy transport.

Guided by these theoretical insights, we fabricated a multilayer superlattice consisting of five periods of alternating 25 nm AlScN and 8 nm AlN layers, with a total film thickness of 190 nm. Cross-sectional transmission electron microscopy (TEM) confirmed the structural periodicity and revealed that AlN interlayers effectively interrupt vertical grain boundary propagation, preventing their traversal through the full film thickness (Figure 2c). High-angle annular dark-field scanning (HADDF) TEM images of AlScN/AlN/AlScN segments (Figure 2d) showed a well-aligned quasi-single-crystalline structure, with interplanar spacings of 2.49 Å and 2.53 Å, corresponding to AlN and AlScN, respectively, consistent with the expected wurtzite crystal parameters[42,43]. These structural features were further corroborated by high-resolution X-ray diffraction (HRXRD) (Extended Data Figure 2). Ferroelectric measurements confirmed robust switching behavior, with a $P_r$ of ~160 μC/cm$^2$ and an increased coercive field of 5.3 MV/cm (Figure 2e). The elevated $E_c$ is attributed to electric field partitioning across the insulating AlN interlayers.

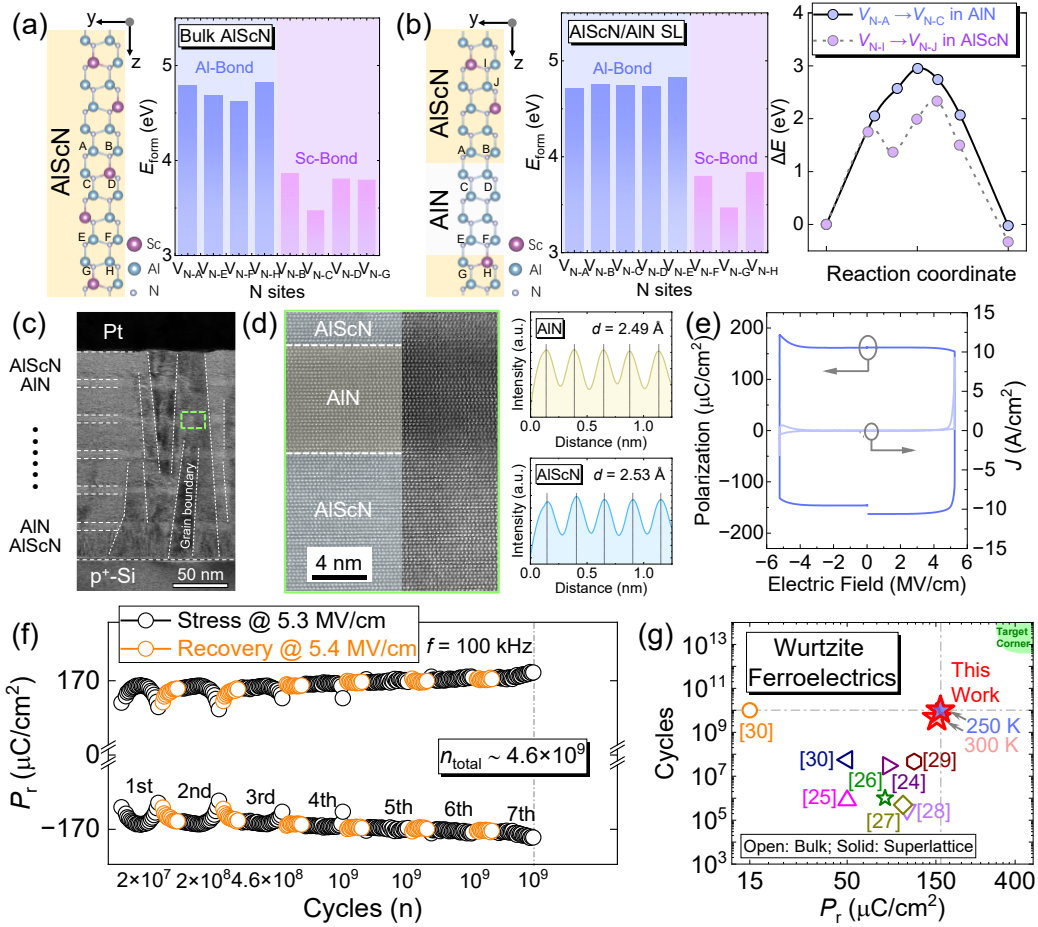

**Figure 2 Wurtzite AlScN/AlN superlattice enables vacancy confinement and endurance breakthrough beyond $10^{10}$ cycles. a-b,** DFT calculations reveal that AlN layers possess significantly higher $E_{form}$ and migration barriers than Sc-rich AlScN regions, serving as effective barriers to vertical $V_N$ transport. **c-e,** TEM and ferroelectric measurements confirm the periodic stacking, quasi-single-crystalline lattice alignment and robust polarization switching. The AlN interlayers notably interrupt vertical grain boundary propagation. **f-g,** Endurance test and benchmark demonstrate a milestone: the first wurtzite ferroelectric superlattice achieves $4.6 \times 10^9$ cycles at 300 K and exceeds $10^{10}$ cycles at 250 K at $2P_r \geq 200\ \mu C/cm^2$, representing a two-order-of-magnitude improvement over previous records[24-30].

Endurance test revealed a substantial enhancement over bulk AlScN, with the superlattice sustaining more than $2\times10^7$ cycles in the first stress epoch, exceeding the breakdown-limited lifetime of bulk counterparts by over an order of magnitude (Figure 2f). Moreover, the dominant failure behavior shifted from abrupt hard breakdown to gradual ferroelectricity loss, consistent with suppressed formation of conductive percolation channels. To further extend device endurance, a recovery operation was introduced after each stress epoch. After seven stress-recovery cycles, the device achieved a cumulative endurance of $4.6\times10^9$ switching cycles at $2P_r \geq 200$ μC/cm$^2$ at room temperature. When tested at a reduced temperature of 250 K, the superlattice exhibited endurance exceeding $1.05\times10^{10}$ cycles under comparable polarization conditions (Extended Data Figure 3). Benchmark against all previously reported wurtzite ferroelectric systems[24-30] (Figure 2g), this work demonstrates the first realization of a superlattice architecture, integrated with a dynamic recovery protocol, that achieves endurance beyond $10^{10}$ cycles while maintaining high remanent polarization[44]. This marks a two-order-of-magnitude improvement over prior art. Taken together, these findings demonstrate that the AlScN/AlN superlattice effectively suppresses vertical vacancy percolation by modulating $E_{form}$ and thus migration barriers. As such, this spatially engineered vacancy confinement strategy provides a viable and CMOS-compatible route to overcome the intrinsic endurance bottleneck of wurtzite ferroelectrics.

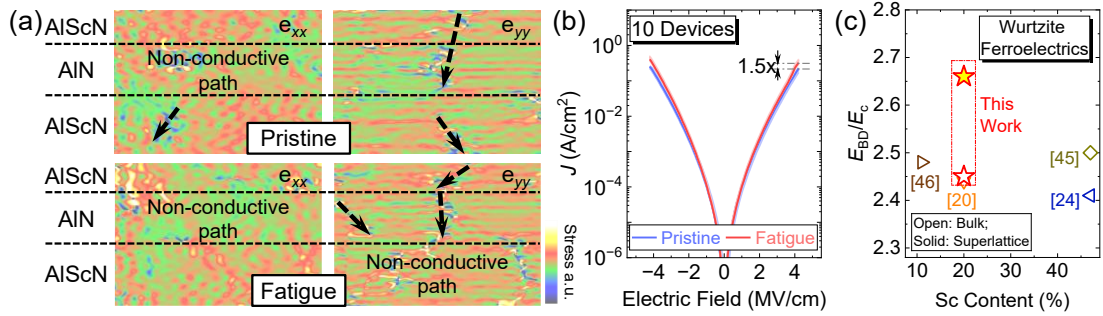

**Figure 3 AlN/AlScN superlattice suppressed dielectric breakdown. a-b,** Strain mapping and leakage measurements reveal minimal conduction path formation and limited current increase after $3\times10^9$ cycles, confirming that AlN interlayers effectively block vertical nitrogen vacancy transport. **b,** $E_{BD}/E_c$ benchmark yields the highest value reported for wurtzite ferroelectrics, underscoring the effectiveness of spatial confinement in preventing defect-driven breakdown[20, 24, 45, 46].

■ **Dual spatial-energetic confinement of nitrogen vacancies underpins the endurance breakthrough**

To elucidate the microscopic origin of the unprecedented endurance observed in AlScN/AlN superlattices, we systematically examined the evolution of lattice structure and nitrogen vacancy dynamics under prolonged cyclic electrical stress. As shown in Figure 3a, strain mapping reveals that even after $3\times10^9$ switching cycles, few continuous conduction paths develop across the film thickness. Vertically aligned grain boundaries are consistently arrested at the AlN interlayers, rather than propagating through the entire stack, confirming that these interlayers act as effective vertical barriers that suppress nitrogen vacancy percolation. This spatial confinement is further corroborated by leakage current measurements: in the superlattice, the median leakage current increases by only ~50% after cycling, in stark contrast to an ~800% rise observed in bulk AlScN after merely $2\times10^6$ cycles (Figure 3b). Consequently, the breakdown-to-coercive field ratio ($E_{BD}/E_c$) reaches a median value of 2.67 across ten devices (Extended Data Figure 4), marking the highest ratio reported to date among wurtzite ferroelectrics with varying scandium concentrations[20, 24, 45, 46] (Figure 3c). These results establish spatially engineered superlattices as a robust platform for suppressing defect-mediated dielectric failure.

Although catastrophic breakdown is effectively mitigated, extended cycling gives rise to a distinct fatigue pathway, manifested as a progressive loss of polarization. To resolve the underlying structural origin, we conducted aberration-corrected transmission electron microscopy (AC-TEM), besides annular bright-field (ABF) imaging, was performed at four representative stages: as-deposited, woken-up, fatigued and recovered (Figures 4a-d). In the pristine state, the wurtzite lattice exhibits a c-axis orientation normal to the growth interface, with polarization aligned downward, commonly referred to as the N-polar state (Figure 4a). Following initial electrical activation, the lattice switches to an upward polarization while preserving crystallographic alignment, corresponding to the M-polar state[5] (Figure 4b). Upon extended cycling, however, localized domains deviate from both canonical states, displaying lattice planes nearly parallel to the growth interface and forming in-plane polarized regions devoid of vertical polarization[47] (Figure 4c and Extended Data Figure 5). Remarkably, these domains can be reverted to the M-polar configuration with recovery operations (Figure 4d), fully restoring out-of-plane polarization. This microstructural evolution directly parallels the macroscopic switching behavior: an initial wake-up phase with increasing remanent polarization, followed by that degradation and eventual recovery[48] .To further identify the defect-driven mechanism responsible for this transformation, nitrogen energy-loss mapping was performed via electron energy-loss spectroscopy (EELS) on identically cycled samples during the first stress epoch (Figures 4e-g). In both the woken-up and recovered states, nitrogen intensity profiles remain spatially uniform and statistically uncorrelated across ten regions, indicative of a homogeneously dispersed vacancy distribution. In stark contrast, the fatigue state exhibits a spatially correlated intensity landscape: clusters of localized adjacent regions show significantly suppressed nitrogen signals relative to their surroundings, providing direct evidence of stress-induced nitrogen vacancy aggregation confined within these localized domains. This redistribution is quantitatively captured by the STD of nitrogen intensity, which increases from 0.030 (woken-up) to 0.038 (fatigued) and subsequently decreases to 0.022 following recovery operation. These findings indicate that vacancy clustering induces strong local depolarization fields, which reorient the lattice away from the out-of-plane axis and consequently suppress vertical polarization[49,50]. Notably, recovery pulses can actively drive vacancies toward energetically favorable lattice sites, dissolving aggregates and restoring out-of-plane ferroelectric order.

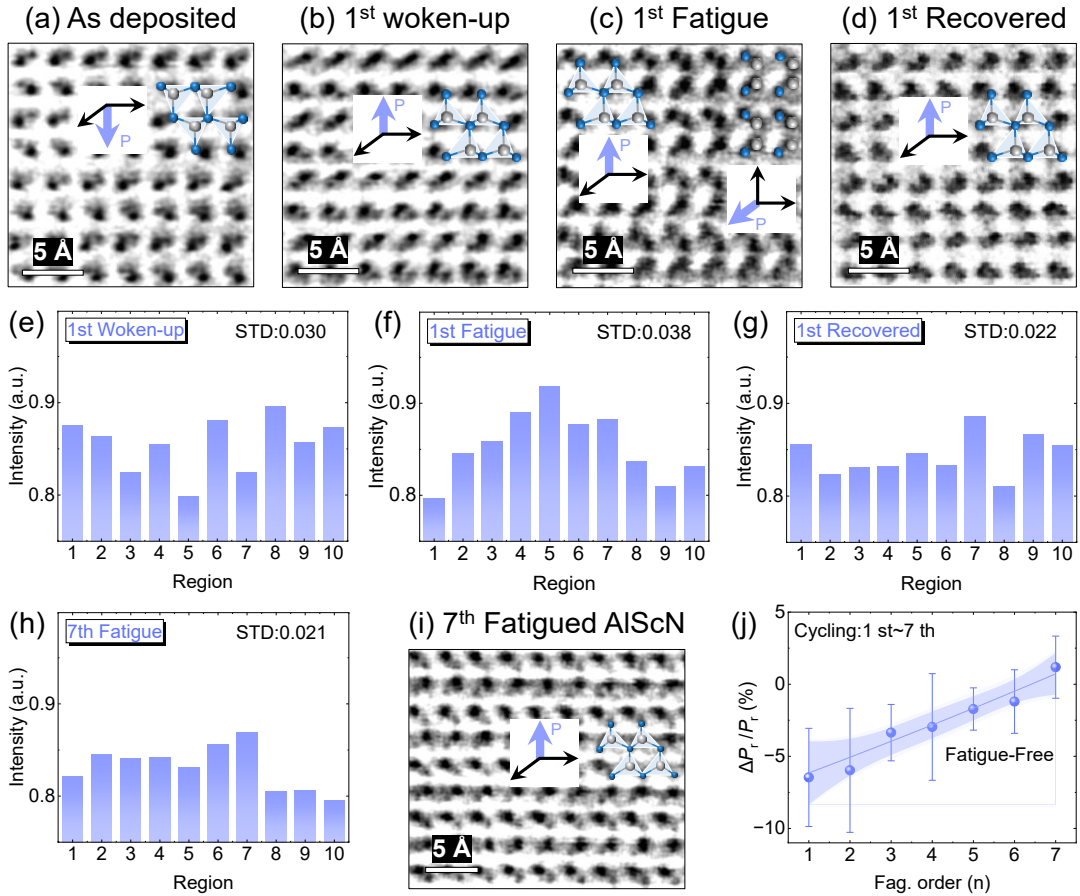

**Figure 4 Vacancy-driven ferroelectricity loss and suppression via stabilized defect topology. a-g,** TEM and EELS reveal that cyclic stress induces nitrogen vacancy clustering and local depolarization fields, reorienting the lattice and suppressing out-of-plane polarization. Recovery pulses dissolve these aggregates by driving vacancies to energetically favorable sites, restoring vertical polarization. **h-j,** In the 7$^{th}$ stress epoch, after $3\times10^9$ cycles, the vacancy distribution remains uniform and the M-polar state is preserved. The sustained reduction in polarization loss across stress-recovery epochs confirms the formation of a self-stabilized vacancy topology via dual spatial-energetic confinement, contributing to the fatigue-resistant ultra-high-cycle operation in wurtzite ferroelectrics.

Figures 4h and 4i present EELS and TEM data acquired after the $7^{th}$ stress epoch, corresponding to over $10^9$ switching cycles. At this advanced cycling stage, the nitrogen vacancy distribution remains remarkably uniform (STD=0.021), while the AlScN lattice consistently retains the M-polar configuration across the entire film. These observations indicate the emergence of a self-stabilized vacancy topology, in which mobile nitrogen vacancies are confined to low-energy immobile configurations that prevent further migration or aggregation. This self-regulated defect landscape preserves both structure and ferroelectricity stability under prolonged electrical stress. As shown in Figure 4j, remanent polarization degradation remains below 3% after $10^9$ cycles, highlighting the pivotal role of dual spatial-energetic confinement in mitigating long-term fatigue and enabling reliable ultra-high-cycle operation in wurtzite ferroelectrics[44].

**Conclusion**

This work identifies the stabilization of nitrogen vacancy topology as the central determinant of endurance in wurtzite ferroelectrics. By combining a spatially engineered AlScN/AlN superlattice with a dynamic recovery protocol, we realize dual spatial-energetic confinement of nitrogen vacancies, enabling reliable ferroelectric switching beyond $10^{10}$ cycles, a two-order-of-magnitude improvement over prior wurtzite systems. This endurance breakthrough represents a critical advance toward the practical deployment of wurtzite ferroelectrics in non-volatile memory technologies. Beyond the record-setting performance, our findings identify nitrogen as the endurance-limiting elemental species in wurtzite systems, providing direct experimental evidence of its governing role in defect regulation and device reliability. Taken together, this work presents a broadly applicable strategy for engineering defect topologies in wurtzite ferroelectrics, enabling CMOS-compatible architectures for next-generation non-volatile memories.

**Methods**

- **Deposition of Bulk AlScN and Wurtzite AlScN/AlN Superlattices and Device Fabrication**

The complete film stack, including bulk AlScN, wurtzite AlScN/AlN superlattices and Pt top electrodes, was deposited in situ in a single uninterrupted vacuum cycle using an SPTS SIGMA 200 sputtering system equipped with five independent targets. Alloyed AlSc (20% Sc), high-purity Al and Pt targets were used to ensure precise control over both chemical composition and interfacial quality. Bulk AlScN films (~190 nm) were deposited from the AlSc target under nitrogen-rich conditions ($N_2$:Ar = 160:32 sccm) at 200 °C using 5 kW DC power. AlN layers were deposited from the Al target under $N_2$:Ar = 95:15 sccm using 6 kW DC power at the same substrate temperature. The superlattice consisted of five bilayers of 25 nm AlScN and 8 nm AlN, followed by a final 25 nm AlScN cap layer, resulting in a total thickness of approximately 190 nm. Pt top electrodes (~50 nm) were subsequently deposited and patterned into circular pads with radii of 10 to 90 μm via standard photolithography and dry etching, forming capacitor structures for subsequent electrical testing. The full fabrication process is illustrated in Extended Data Figure 6.

■ **Basic electrical characterization**

Basic electrical measurements on both bulk AlScN films and wurtzite AlScN/AlN superlattices included polarization switching, leakage current, capacitance and dielectric breakdown characterization. All measurements were performed using a Lake Shore CRX-4K cryogenic probe station. Polarization was measured using the positive-up negative-down (PUND) method on an aixACCT TF 3000 system. Each pulse sequence comprised five rectangular voltage pulses: one negative preset pulse, followed by two positive and two negative pulses. The pulse width and rise time were set to 500 μs and 250 μs, respectively. The pulse amplitude was fixed at 4.7 MV/cm for bulk AlScN and 5.3 MV/cm for AlScN/AlN superlattices. Leakage current and breakdown tests were conducted using current-voltage (*I-V*) and time-zero dielectric breakdown (TZDB) protocols on a Keithley 4200A-SCS parameter analyzer. Capacitance-voltage (*C-V*) measurements were carried out in a quasi-static small-signal regime with a 100 kHz AC excitation of 0.03 MV/cm.

◼ **Endurance characterization and recovery operation**

Endurance tests were conducted by applying continuous bipolar triangular voltage pulses using the pulse generator integrated into the TF 3000 system. Each cycle consisted of one complete positive and negative switching event, applied at a repetition rate of 100 kHz. PUND measurements were periodically inserted to monitor the evolution of $P_r$ as a function of cycle count. The pulse amplitude was maintained at 4.7 MV/cm for bulk AlScN and 5.3 MV/cm for AlScN/AlN superlattices unless otherwise stated. All endurance measurements were conducted on the Lake Shore CRX-4K probe station, with sample temperature stabilized at either 300 K or 250 K depending on test conditions. Recovery operations were applied prior to catastrophic failure, as guided by behavior observed in parallel-stressed reference devices. Recovery was implemented by introducing 1,000 additional cycles of bipolar triangular pulses under identical conditions (100 kHz), using the same instrumentation.

◼ **Cs-TEM characterization and EELS Analysis**

Cross-sectional TEM samples were prepared using a Helios G5 CX focused ion beam system through mechanical thinning followed by $Ga^+$ ion milling. High-angle annular dark-field (HAADF) and annular bright-field (ABF) scanning transmission electron microscopy (STEM) images were acquired using a double aberration-corrected electron microscope (JEM-ARM300F2). Image processing was performed using the HREM Filter software (HREM Research Inc.) to reduce background noise while preserving atomic-resolution features. Electron energy-loss spectroscopy (EELS) spectra were collected in STEM mode of a double aberration-corrected electron microscope (JEM-ARM300F2) with an aperture of 5 mm and an energy dispersion of 90 meV per channel. Strain mapping was performed using geometric phase analysis (GPA) on HAADF-STEM images, using a Digital Micrograph-compatible plugin.

■ **First-Principles Calculations**

Density functional theory (DFT) calculations were carried out using the QUANTUM ESPRESSO package (v7.0). Ultrasoft pseudopotentials (USPP) were employed within the generalized gradient approximation (GGA) using the Perdew-Burke-Ernzerhof (PBE) exchange-correlation functional. Extended Data Figure 7 shows the constructed supercells for bulk AlScN and the wurtzite AlScN/AlN superlattice. The bulk AlScN supercell was based on a 1×2×14 expansion of the wurtzite AlN primitive cell (space group P6$_3$mc; a = b = 3.129 Å, c = 5.017 Å). Scandium doping was introduced by partially substituting Al atoms at selected lattice sites to obtain the target composition. The superlattice was generated by periodically replacing specific AlScN layers in the bulk model with undoped AlN layers, forming a vertically stacked structure along the [0001] direction with a thickness ratio of 5:2.

The nitrogen vacancy formation energy ($E_{form}$) was calculated using the expression:

$$E_{form} = E_0^{N-1} + \frac{1}{2}E_{N_2} - E^N$$

where $E_0^{N-1}$ and $E^N$ represent the total energies of the supercell with and without a nitrogen vacancy, respectively, and $E_{N_2}$ is the total energy of an isolated N$_2$ molecule in the gas phase. Migration barriers for nitrogen vacancies were evaluated using the nudged elastic band (NEB) method.

**Acknowledgements**

This work was supported by the National Science and Technology Major Project (No. 2022ZD0119002), the National Natural Science Foundation of China (Grant No. 62025402, 92264101), the Major Program of Zhejiang Natural Science Foundation (Grant No. LD25F040004), and the Fundamental Research Funds for the Central Universities (No. YJSJ25013).

**Author Contributions**

G. Han supervised the project. J. Zhou and G. Han conceived the research, designed the experiments, and led the data interpretation and discussion. R. Wang fabricated all samples and performed the electrical measurements. F. Zhu and R. Wang conducted the TEM characterizations and EELS analyses. Q. Hao and J. Chen carried out the first-principles calculations and interpreted the computational results. S. Zheng and W. Sun developed the nitride deposition protocols. J. Zhou and R. Wang co-wrote the manuscript with input from all authors. All authors contributed to data analysis and scientific discussions.

**Competing interests.**

The authors declare no competing financial interest.

**Extended Data Figures.**

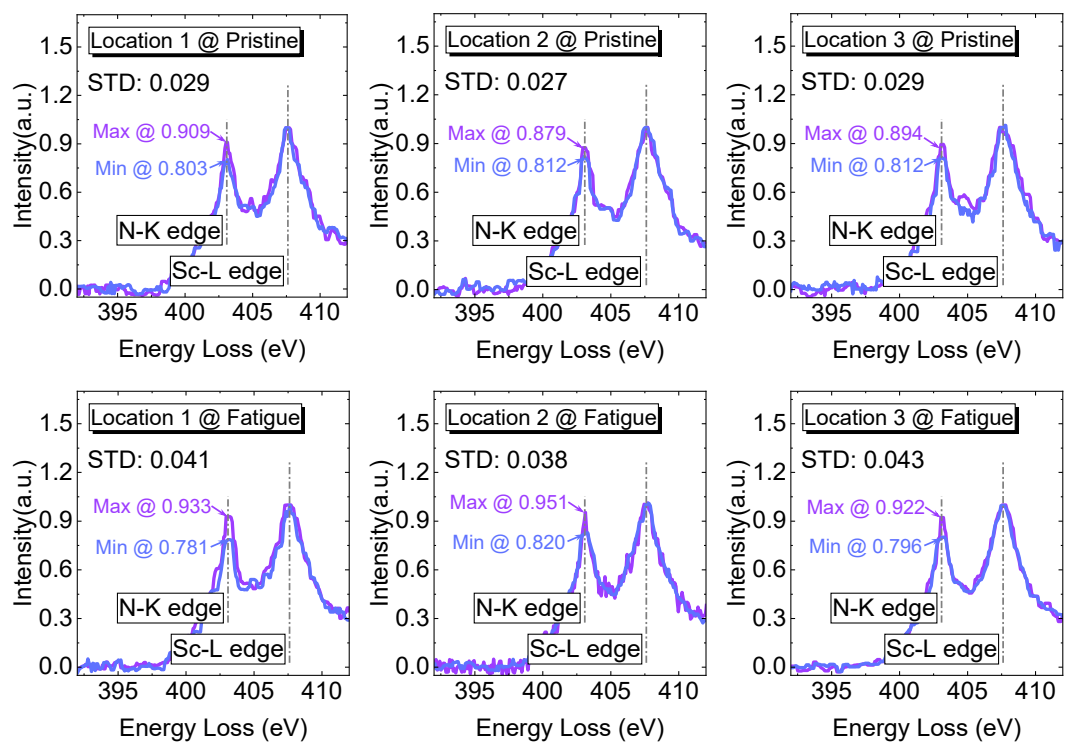

**Extended data Figure 1. Raw EELS spectra supporting vacancy clustering analysis in bulk AlScN.** EELS spectra spanning 392-412 eV, covering both the N-K and Sc-L edges, were acquired from ten distinct regions across three lateral locations in pristine and fatigued bulk AlScN films. The N-K edge intensity serves as a proxy for nitrogen vacancy concentration, while the Sc-L edge, originating from the fixed Sc content, provides an internal normalization reference. Although only representative spectra corresponding to the maximum and minimum N-K edge intensities are displayed here, the full dataset forms the basis for Figure 1e. These measurements reveal lateral redistribution and directional clustering of nitrogen vacancies under cyclic electrical stress, supporting the correlation between vacancy migration and dielectric degradation pathways.

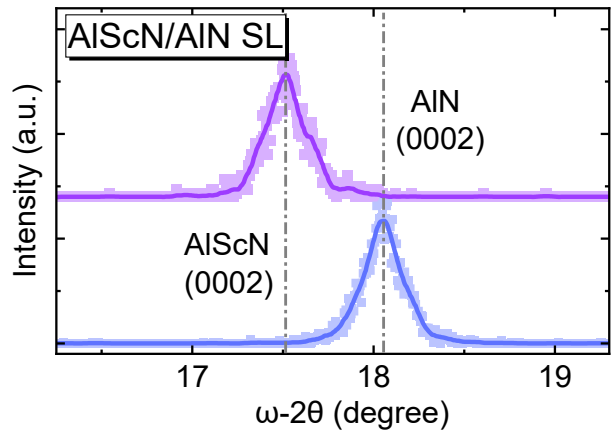

**Extended Data Figure 2. X-ray diffraction evidence of structural periodicity in AlScN/AlN superlattices.** High-resolution ω-2θ X-ray diffraction was performed to resolve the layered structure of the AlScN/AlN superlattice. The diffraction profile near 18° reveals two distinct peaks corresponding to the (0002) planes of AlScN and AlN, located at ~17.5° and ~18.1°, respectively. The peak separation reflects the difference in interplanar spacing between the two wurtzite phases, confirming the successful formation of a periodic multilayer stack. This result supports the TEM observations of well-defined AlScN/AlN interfaces and validates the designed superlattice geometry, in which AlN layers interrupt vertical grain boundary propagation.

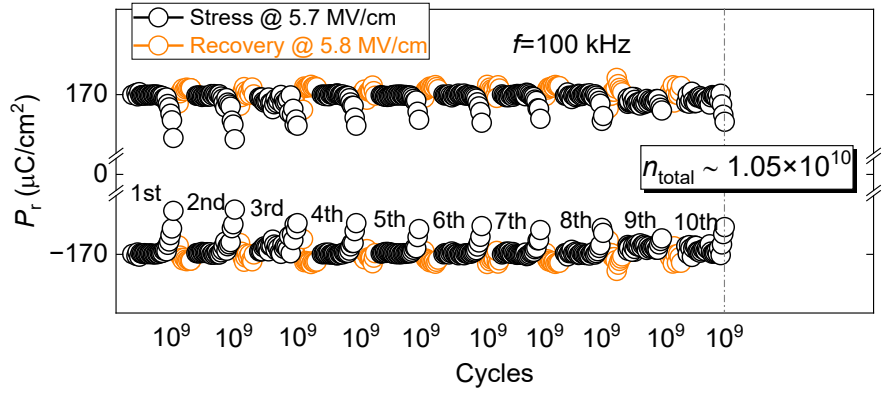

**Extended Data Figure 3. Endurance beyond $10^{10}$ cycles in AlScN/AlN superlattices at 250 K.** Cumulative fatigue testing was performed at 250 K under ten consecutive stress-recovery epochs. The AlScN/AlN superlattice sustained robust ferroelectric switching with $2P_r \geq 200$ μC/cm$^2$, achieving over $1.05 \times 10^{10}$ cycles without failure.

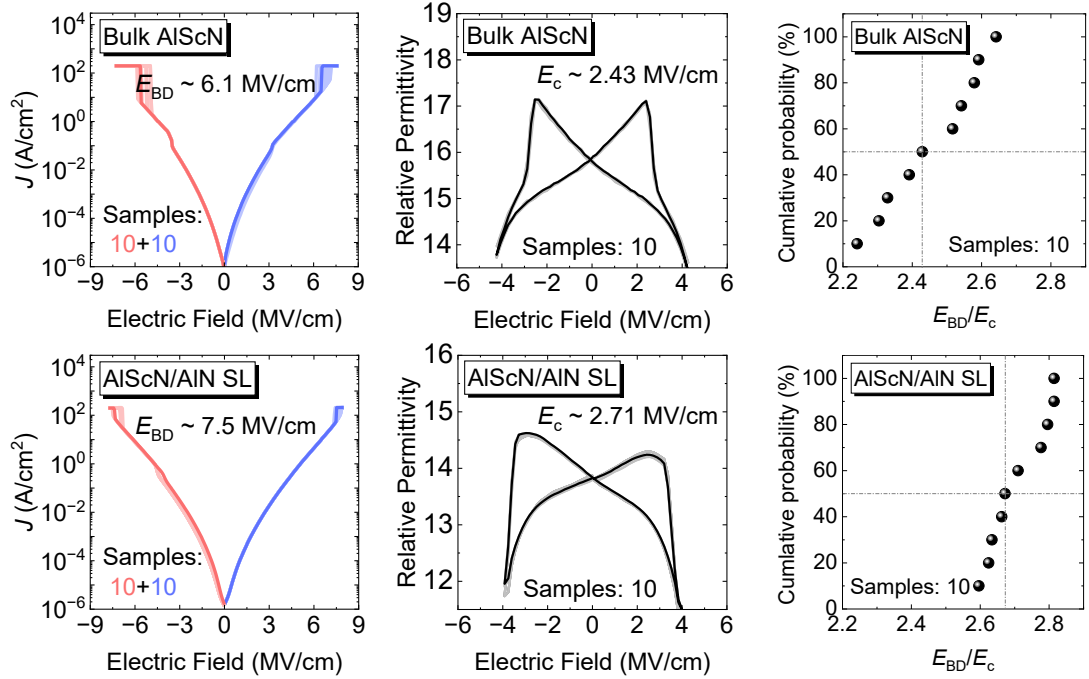

**Extended Data Figure 4. Enhanced breakdown-to-coercive field ratio in AlScN/AlN superlattices.** Time-zero dielectric breakdown (TZDB) and quasi-static *C-V* measurements were conducted on both bulk AlScN and AlScN/AlN superlattices to extract intrinsic $E_{BD}$ and $E_c$ values. The AlScN/AlN superlattice, benefiting from embedded AlN interlayers, exhibits a significantly elevated $E_{BD}/E_c$ ratio, reaching a median of 2.67 across ten devices. This marked improvement confirms the role of spatial vacancy confinement in suppressing defect-induced dielectric failure and establishes a new endurance benchmark among wurtzite ferroelectrics.

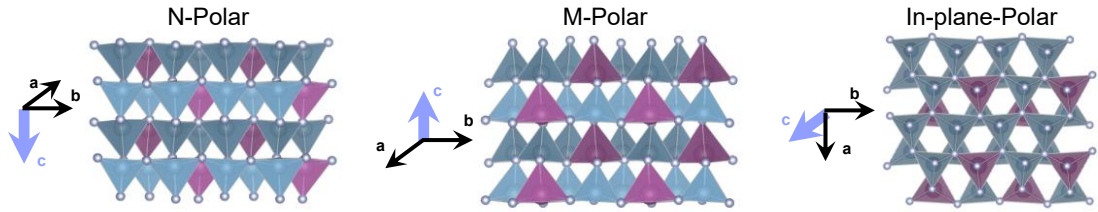

**Extended Data Figure 5. Three representative crystal structures of wurtzite AlScN under electrical stress.** Schematic models depict the three polarization states observed in AlScN during electrical stress: the N-polar state with downward polarization along the c-axis (as-deposited), the M-polar state with upward polarization (after wake-up), and the in-plane polarized state featuring reoriented lattice planes nearly parallel to the growth interface and suppressed out-of-plane polarization.

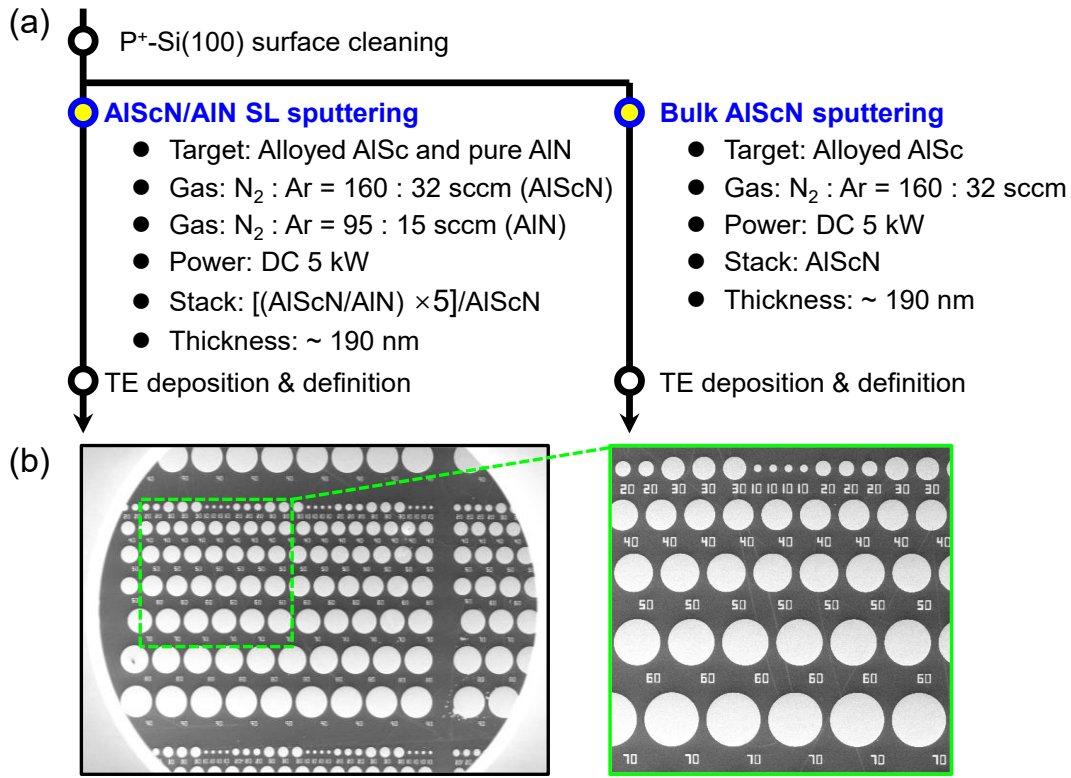

**Extended Data Figure 6. Fabrication process and top-view morphology of bulk AlScN and AlScN/AlN superlattice capacitors. a,** Schematic process flow for the deposition of bulk AlScN and AlScN/AlN superlattices, followed by Pt electrode patterning to form capacitors. All layers were deposited in situ using an SPTS SIGMA 200 sputtering system within a single uninterrupted vacuum cycle. **b,** Top-view SEM image of a fabricated capacitor, showing patterned circular Pt electrodes with radius ranging from 10 to 90 μm.

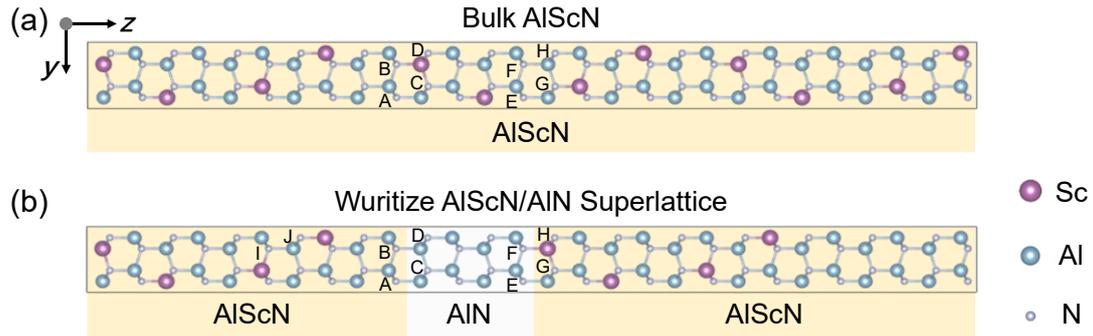

**Extended Data Figure 7. Atomistic models of bulk AlScN and AlScN/AlN superlattices used for vacancy energetics calculations. a,** Supercell structure of bulk AlScN, constructed by a 1×2×14 expansion of the wurtzite AlN primitive cell (space group P6$_3$mc; a = b = 3.129 Å, c = 5.017 Å), with partial substitution of Al atoms by Sc to achieve the target composition. **b,** Supercell structure of the AlScN/AlN superlattice, generated by periodically replacing selected AlScN layers with undoped AlN, forming a vertically stacked configuration along the [0001] direction with a thickness ratio of 5:2. These models serve as the basis for first-principles calculations of nitrogen vacancy formation energies and spatial confinement effects.